\documentclass[iop]{emulateapj}
\usepackage{graphicx}
\usepackage{epsfig}
\usepackage{aas_macros}

\begin{document}

\title{Implications from the upper limit of radio afterglow emission of FRB 131104/Swift J0644.5-5111}

\author{He Gao$^{1}$, and Bing Zhang$^{2,3,4}$}
\affil{$^1$Department of Astronomy, Beijing Normal University, Beijing 100875, China; gaohe@bnu.edu.cn\\
  $^2$Department of Physics and Astronomy, University of Nevada Las Vegas, NV 89154, USA;\\
$^3$Department of Astronomy, School of Physics, Peking University, Beijing 100871, China; \\
$^4$Kavli Institute of Astronomy and Astrophysics, Peking University, Beijing 100871, China;\\}

\begin{abstract}
A $\gamma$-ray transient, Swift J0644.5-5111, has been claimed to be associated with FRB 131104. 
However, a long-term radio imaging follow-up observations only placed an upper limit on the radio afterglow flux of Swift J0644.5-5111. Applying the external shock model, we perform a detailed constraint on the afterglow parameters for the FRB 131104/Swift J0644.5-5111 system. We find that for the commonly used microphysics shock parameters (e.g., $\epsilon_e=0.1$, $\epsilon_B=0.01$ and $p=2.3$), if the FRB is indeed cosmological as inferred from its measured dispersion measure (DM),
the ambient medium number density should be $\leq 10^{-3}~\rm{cm^{-3}}$, which is the typical value for a compact binary merger environment but disfavors a massive star origin. Assuming a typical ISM density, one would require that the redshift of the FRB much smaller than the value inferred from DM ($z\ll0.1$), implying a non-cosmological origin of DM. The constraints are much looser if one adopts smaller $\epsilon_B$ and $\epsilon_e$ values, as observed in some GRB afterglows. The FRB 131104/Swift J0644.5-5111 association remains plausible. We critically discuss possible progenitor models for the system.
\end{abstract}

\section{Introduction}

Recently, a new type of millisecond radio burst transients, named fast radio bursts (FRBs), has attracted wide attention \citep{lorimer07,thornton13}. Due to limited data, especially the lack of distance measurements, their physical origin is subject to intense debate. In the literature, 18 FRBs have been reported up to now, and their observed event rate is predicted to be $\sim 10^{-3}$
$\rm gal^{-1}$ $\rm yr^{-1}$. Most of these bursts are located at high galactic latitudes and have anomalously large
dispersion measures (DM), which strongly suggest that they are of an extragalactic or even cosmological origin. Attributing DM to an intergalactic medium origin, the corresponding redshifts $z$ are around 0.5-1. The isotropic total energy released in a FRB is $\sim (10^{38}-10^{40})$ erg, and the peak radio luminosity is $\sim (10^{42}-10^{43})$ erg $\rm s^{-1}$ \citep{thornton13}.

Many models have been proposed to account for FRBs. These include non-catastrophic events such as magnetar giant flares \citep{popov10,kulkarni14,katz16b}, giant pulses of young pulsars \citep{cordes16,connor16}, galactic flaring stars \citep{loeb14}, asteroid collisions with neutron stars \citep{geng15,dai16a},  and catastrophic events such as blitzars \citep{falcke14,zhang14}, neutron star mergers \citep{totani13,zhang14,wang16}, white dwarf mergers \citep{kashiyama13}, charged black hole mergers \citep{zhang16a,liu16}, etc.

Considering the dramatic observational event rate, once their cosmological origin could be confirmed, FRBs have great potential to become a powerful cosmological probe. For instance, they could be used to identify missing baryons \citep{mcquinn14}, infer the baryon content and reionization history of the universe \citep{deng14,zheng14,keane16}, constrain cosmological parameters and the dark matter equation of state \citep{gao14,zhou14}, conduct fundamental tests of Einstein's Weak Equivalent Principle \citep{wei15}, and even set a stringent upper limit on the mass of photons \citep{wu16}. 

Some interesting observational features have been collected recently for individual FRBs. FRB 140514 was the first real-time FRB detection and the first with polarization information, i.e. a $21\pm7\%~(3\sigma)$ circular polarization on the leading edge and a $1\sigma$ upper limit ($<10\%$) on linear polarization. Later, an examination of archival data revealed Faraday rotation in FRB 110523 \citep{petroff15}. The $44\%$ linear polarization of FRB 110523 enabled the first simultaneous measurements of both its dispersion measure and rotation measure \citep{masui15}. FRB 121002 was reported to exhibit a double-peaked profile, with two peaks separated by 5ms \citep{champion16}. \cite{spitler16} reported the detections of 10 additional bright bursts from FRB 121102 at 1.4 GHz at the Arecibo Observatory, making FRB 121102 the first and the only one so far with repeating features. \cite{keane16} claimed to find a bright radio fading transient following FRB 150418, which lasted for about 6 days. A putative host galaxy with redshift $0.492\pm0.008$ was identified. However, subsequent observations suggested that the counterpart radio transient is more likely  attributed to AGN variability instead of an afterglow of FRB 150418 \citep{williams16}. 

The possibility that a small fraction of FRBs could be associated with gamma-ray bursts (GRBs) was proposed by \cite{zhang14}. Most recently, a $\gamma$-ray transient, Swift J0644.5-5111, was claimed to be associated with FRB 131104 both spatially and temporarily \citep{delaunay16}. 
The $\gamma$-ray fluence of the transient is 10 order of magnitude larger than the fluence of FRB 131104, and
the total $\gamma$-ray energy output is $E_\gamma \approx 5\times 10^{51}$\,erg at the nominal redshift $z\approx 0.55$ inferred from the dispersion measure of FRB 131104 \citep{delaunay16}\footnote{With a more rigorous formula \citep{deng14,gao14}, the inferred redshift is $z \sim 0.7$ if one assumes that the host galaxy DM is $\sim 100 ~{\rm cm^{-3}~pc}$, which corresponds to a $\gamma$-ray energy of $8.3 \times 10^{51}~{\rm erg}$.}.  X-ray, ultraviolet, and optical observations starting from two days after the FRB did not reveal any associated afterglow. \cite{murase16} discussed the general implications of bright $\gamma$-ray counterpart to the FRB, and suggested that, largely independent of the source model, a radio afterglow is expected for such FRB/GRB systems. Soon after, \cite{shannon16} reported their radio imaging follow-up observations of the original localization region of the FRB, beginning from three days after the event and comprising 25 epochs over 2.5~yr. They reported a coincident AGN flare similar to the radio transient observed in the error box of FRB 131104 (but with a smaller amplitude), but no radio afterglow was detected within the Swift J0644.5-5111 error circle. An upper limit of 70 $\mu \rm Jy$ and 100 $\mu \rm Jy$ were placed on the radio afterglow emission of FRB/GRB system at 5.5 GHz and 7.5 GHz, respectively.

Here we present a comprehensive discussion for the physical implications of the null detection of radio afterglow for FRB 131104/Swift J0644.5-5111. Throughout the paper, the convention $Q = 10^nQ_n$ is adopted in cgs units.

\section{FRB 131104/Swift J0644.5-5111}

FRB 131104 is a real-time discovery detected in a targeted observation of the Milky Way satellite Car dSph (with Galactic latitude $b=22.2^{\circ}$) by the Parkes radio telescope \citep{ravi15}. The event was detected at 18:04:01.2 on UT 2013~November~4 (MJD~56600), in beam 5 of the MB receiver, which at the time was pointed at the celestial coordinates (J2000) 06h44m10.4s, $-$51d16m40s. The DM of the burst is 779\,cm$^{-3}$\,pc, which exceeds the prediction for the maximum line-of-sight Galactic contribution by a factor of 11. The redshift of FRB 131104 inferred from its DM value is $0.55$. 

Swift J0644.5-5111 is an un-triggered $\gamma$-ray transient identified through a dedicated Swift/BAT sub-threshold search for the $\gamma$-ray counterpart of FRB 131104. The transient position is at R.A. $={06}^{h}{44}^{m}{33}^{s}.{12}$, Dec. $=-51^{\circ}11'31".2$ (J2000), with $r_{90}=6\farcm 8$. It is located near the edge of the BAT field of view, with only 2.9\% of BAT detectors illuminated through the coded mask \citep{delaunay16}. The 15--150\,keV discovery image started from UTC 18:03:52, with a point-source-like excess $\gamma$-ray transient peaking at a signal-to-noise ratio $\mathcal{S} = 4.2\sigma$. The transient has a duration $T_{90} \approx 377$\,s. The $\gamma$-ray fluence of the transient is $S_\gamma\approx 4\times 10^{-6}~\rm{erg~cm^{-2}}$, about 10 orders of magnitude larger than the fluence of FRB 131104, $S_{\rm FRB}  \approx 3\times 10^{-16}~\rm{erg~cm^{-2}}$ \citep{ravi15}. With the inferred redshift $0.55$, the $\gamma$-ray energy output is $E_\gamma \approx 5\times 10^{51}$\,erg.

\section{Afterglow model}

Any FRB may leave behind an afterglow, the brightness of which depends on the radio emission efficiency \citep{yi14}. The lower the efficiency, the brighter the afterglow. If Swift J0644.5-5111 is associated with the FRB, such an efficiency is settled ($\sim 10^{-10}$), which suggests a bright cosmic fireball. The absolute energy of the fireball depends on the distance to the source. For the nominal distance $z=0.55$, the total isotropic energy falls into the range of both long and short GRBs. Assuming a relativistic fireball similar to cosmological GRBs, the deceleration of the fireball would power a radio afterglow that may be detectable \citep{murase16}.

Assuming that a power-law spectrum of the accelerated electrons is a power law function with the index of $p$, and that a constant fraction $\epsilon_e$ of the shock energy is distributed to electrons, the minimum injected electron Lorentz factor could be estimated as (for $p>2$)
\begin{eqnarray}
\gamma_m=\frac{(p-2)}{(p-1)}\epsilon_e(\gamma -1) \frac{m_p}{m_e},
\end{eqnarray}
where  $m_p$ and $m_e$ is proton mass and electron mass respectively. Assuming that the magnetic energy density behind
the shock is a constant fraction $\epsilon_B$ of the shock energy density, one can obtain
$B'=(8 \pi e\epsilon_B)^{1/2}$,
where $e$ is the energy density in the shocked region.

According to standard synchrotron radiation model, the observed radiation power and the characteristic frequency of an electron with Lorentz factor $\gamma_e$ are given by
$\label{power} P(\gamma_e) \simeq \frac 4 3 \sigma_T c \gamma^2
\gamma_e^2 \frac {B'^2} {8\pi}$,
$\label{freq} \nu(\gamma_e) \simeq \gamma \gamma_e^2 \frac {q_e B'}
{2 \pi m_e c}$,
where the factors of $\gamma^2$ and $\gamma$ are introduced to transform the values from the rest frame of the shocked fluid to the frame of the observer. The spectral power would peak at $\nu(\gamma_e)$, with an approximate value
\begin{eqnarray}
\label{flux} P_{\nu,\rm max}\approx\frac{P(\gamma_e)}{\nu(\gamma_e)}=
\frac {m_e c^2 \sigma_T} {3 q_e} \gamma B'~.
\end{eqnarray}
The spectra would vary as $\nu^{1/3}$ for $\nu<\nu(\gamma_e)$, and cuts off essentially exponentially for
$\nu>\nu(\gamma_e)$. 

Considering the cooling of radiated electrons,  a critical electron Lorentz factor $\gamma_c$ could be defined by setting $\tau(\gamma_e)=t$, where 
$\tau(\gamma_e)={(\gamma\gamma_e m_e c^2)}/{(\frac 4 3 \sigma_T c
\gamma^2 \gamma_e^2 \frac {B'^2} {8\pi})}={(6\pi m_e
c)}/{(\gamma\gamma_e \sigma_T B'^2) }$
is the life time of a relativistic electron with Lorentz factor $\gamma_e$ in the observer frame and $t$ refers to the dynamical timescale in the observer frame. One can then define a ``cooling" Lorentz factor
\begin{eqnarray}
\label{cool} \gamma_c= \frac{6\pi m_e c}{\gamma \sigma_T B'^2t},
\end{eqnarray}
above which electrons lose most of their energies within the dynamical time scale,
so that the electron distribution shape is modified in the $\gamma_e>\gamma_c$ regime.

There is a third characteristic frequency $\nu_a$, below which synchrotron self-absorption is important. One may define $\nu_a$ by equating the synchrotron flux and the flux of a blackbody, i.e.
\begin{eqnarray}
I_{\nu}^{syn}(\nu_a)=I_{\nu}^{bb}(\nu_a)\simeq2kT\cdot\frac{\nu_a^2}{c^2}
\end{eqnarray}
where the blackbody temperature is
\begin{eqnarray}
kT={\rm max} [\gamma_{a},{\rm min}(\gamma_c,\gamma_m)] m_e c^2,
\end{eqnarray}
and $\gamma_{a}$ is the corresponding electron Lorentz factor of $\nu_a$ for synchrotron radiation, i.e. $\gamma_a=(4\pi m_e c\nu_a/3eB')^{1/2}$ \citep[e.g.][]{kobayashizhang03}.

The afterglow lightcurve for any frequency can be calculated once the blastwave dynamics (i.e. $\gamma$ as a function of $R$ or $t$) is specified. We consider a constant density ambient medium and $t>3$ days during which 
the blastwave has entered the Blandford-McKee self-similar deceleration regime \cite[see][for a review]{gao13}. 
In this regime, one can write the characteristic frequencies ($\nu_m$, $\nu_c$, $\nu_a$) and the peak synchrotron flux density $F_{\rm{\nu,max}}$ as\footnote{Here we assume that all the electrons in the shock are accelerated, i.e., the injection fraction of non-thermal electrons $f_e$ is around unity. The expressions for $\nu_m$, $\nu_a$ and $F_{\nu,\rm max}$ would be modified if $f_e \neq 1$, and the constraints on the number density would become looser if $f_e \ll 1$ (see \citealt{murase16} for analytical expressions and detailed discussion about the effect of $f_e$).} \citep{meszaros97,sari98,yost03,gao13}.
\begin{eqnarray}
&&\nu_m=   3.8\times10^{11}~{\rm Hz}~(1+z)^{1/2}\frac{G(p)}{G(2.3)}E_{52}^{1/2}\epsilon_{e,-1}^{2}\epsilon_{B,-2}^{1/2}t_{\rm day}^{-3/2}           ,\nonumber\\
&&\nu_c=    3.1\times10^{16}~{\rm Hz}~(1+z)^{-1/2}E_{52}^{-1/2}n_{0,0}^{-1}\epsilon_{B,-2}^{-3/2}t_{\rm day}^{-1/2}           \nonumber\\
&&F_{\rm{\nu,max}}=     2.3\times10^{3}~\mu {\rm
Jy}~(1+z) E_{52}^{}n_{0,0}^{1/2}\epsilon_{B,-2}^{1/2}D_{28}^{-2}          ,\nonumber\\
&&\nu_a=5.7\times10^{9}~{\rm Hz}~(1+z)^{-1}\frac{g^{\rm I}(p)}{g^{\rm I}(2.3)}E_{52}^{1/5}n_{0,0}^{3/5}\epsilon_{e,-1}^{-1}\epsilon_{B,-2}^{1/5},\nonumber\\
&&~~~~~~~~~~~~~~~~~~~~~~~~~~~~~~~~~~~~~~~~~~~~~~~~~\nu_a < \nu_m < \nu_c\nonumber \\
&&\nu_a=1.7\times10^{10}~{\rm Hz}~(1+z)^{\frac{p-6}{2(p+4)}}\frac{g^{\rm II}(p)}{g^{\rm II}(2.3)}E_{52}^{\frac{p+2}{2(p+4)}}n_{0,0}^{\frac{2}{p+4}}\nonumber\\
&&~~~~~~~\epsilon_{e,-1}^{\frac{2(p-1)}{p+4}}\epsilon_{B,-2}^{\frac{p+2}{2(p+4)}}t_{\rm day}^{-\frac{3p+2}{2(p+4)}},
~~~~~~~~~~~~~~\nu_m < \nu_a < \nu_c\nonumber \\
\end{eqnarray}
where 
$G(p) = [(p-2)/(p-2)]^2$, $g^{\rm I}(p)= \frac{(p-1)}{(p-2)} (p+1)^{3/5} f(p)^{3/5}$, 
$g^{\rm II}(p)= e^{\frac{11}{p+4}}\left(\frac{p-2}{p-1}\right)^{\frac{2(p-1)}{p+4}}(p+1)^{\frac{2}{p+4}}f(p)^{\frac{2}{p+4}}$,
and $f(p)=\left[{\Gamma(\frac{3p+22}{12})\Gamma(\frac{3p+2}{12})}\right]/\left[{\Gamma(\frac{3p+19}{12})\Gamma(\frac{3p-1}{12})}\right]$. 

For the time scale we are interested in (e.g., $t>3$ days), the external shock emission would be in the ``slow cooling ($\nu_m<\nu_c$) regime for reasonable parameters \citep{sari98}. The observational frequencies (5.5 GHz and 7.5 GHz) in any case are smaller than $\nu_c$. In this case, the observational flux could be estimated as 
\begin{eqnarray}
\label{gp} F_{\nu_{\rm obs}} = F_{\rm{\nu,max}}\left\{ \begin{array}{ll} \left(\frac{\nu_a}{\nu_m}\right)^{1/3}\left(\frac{\nu_{\rm obs}}{\nu_a}\right)^{2}, &\nu_{\rm obs}<\nu_a;\\
\left(\frac{\nu_{\rm obs}}{\nu_m}\right)^{1/3}, &\nu_a<\nu_{\rm obs}<\nu_m; \\
\left(\frac{\nu_{\rm obs}}{\nu_m}\right)^{-(p-1)/2}, &\nu_m<\nu_{\rm obs}<\nu_c; \\
\end{array} \right.
\end{eqnarray}
when $\nu_a<\nu_m<\nu_c$, or 
\begin{eqnarray}
\label{gp} F_{\nu_{\rm obs}} = F_{\rm{\nu,max}}\left\{ \begin{array}{ll} \left(\frac{\nu_m}{\nu_a}\right)^{(p+4)/2}\left(\frac{\nu_{\rm obs}}{\nu_m}\right)^{2}, &\nu_{\rm obs}<\nu_m;\\
\left(\frac{\nu_a}{\nu_m}\right)^{-(p-1)/2}\left(\frac{\nu_{\rm obs}}{\nu_a}\right)^{5/2}, &\nu_m<\nu_{\rm obs}<\nu_a; \\
\left(\frac{\nu_{\rm obs}}{\nu_m}\right)^{-(p-1)/2}, &\nu_a<\nu_{\rm obs}<\nu_c; \\
\end{array} \right.
\end{eqnarray}
when  $\nu_m<\nu_a<\nu_c$. 

\begin{figure*}[t]
\begin{center}
\begin{tabular}{lll}
\resizebox{85mm}{!}{\includegraphics[]{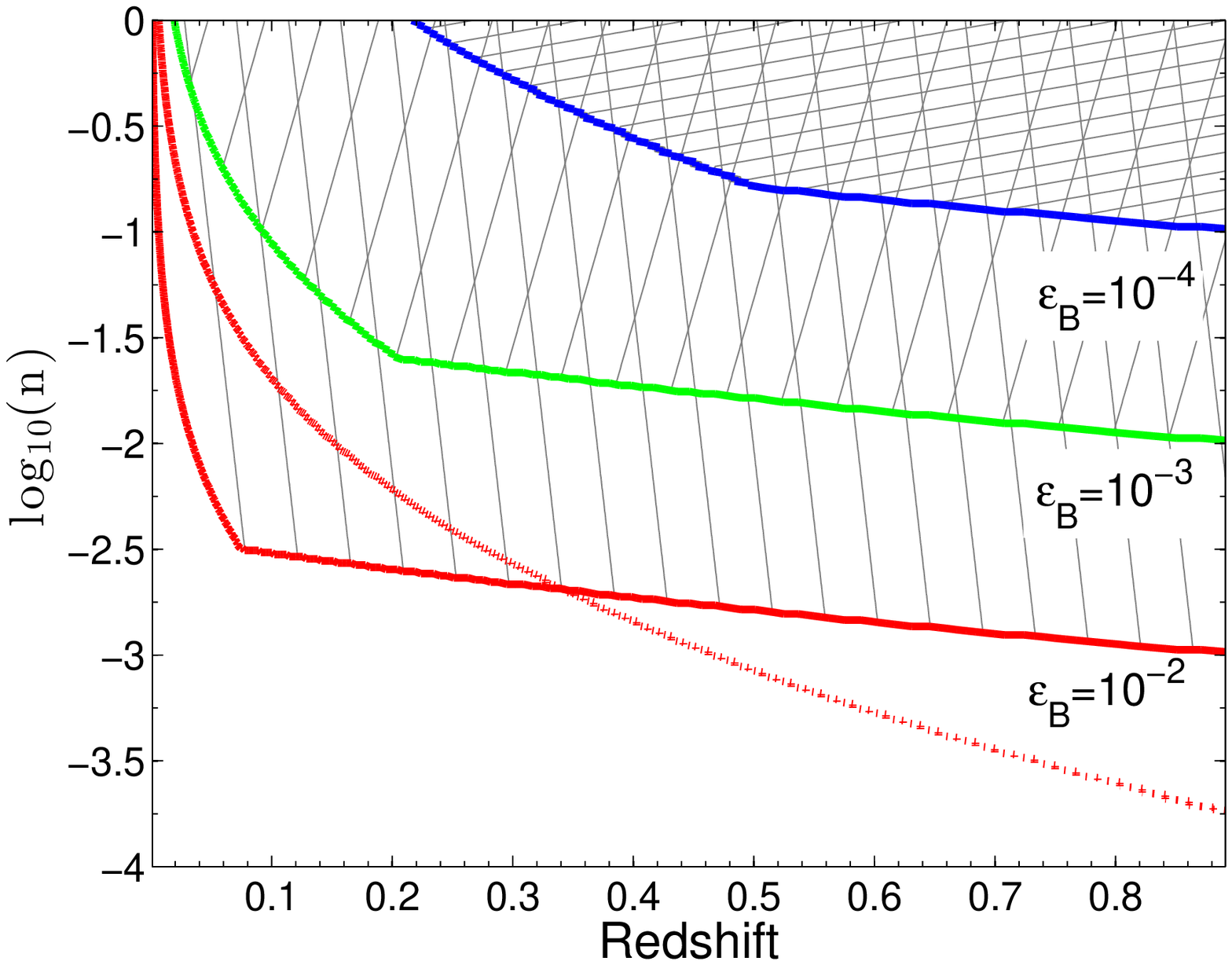}} &
\resizebox{85mm}{!}{\includegraphics[]{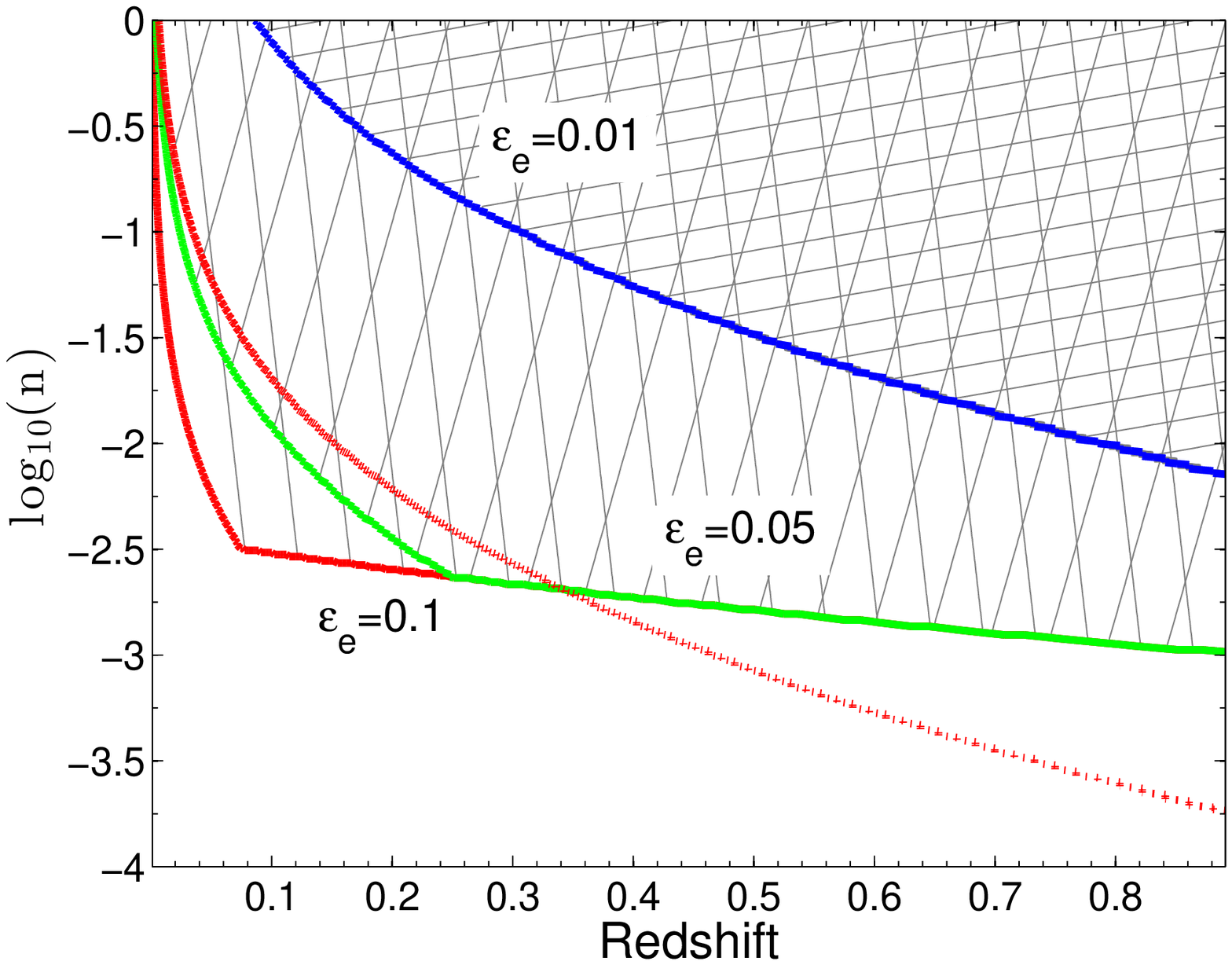}} &
\end{tabular}
\caption{The rejection regions (shadowed) in the 2D $(z,n)$ space based on the upper limit of the radio afterglow emission, 70$\mu$Jy at 5.5 GHz. The left panel fixes $\epsilon_e = 0.1$ and varies $\epsilon_B$, while the right panel fixes $\epsilon_B = 0.01$ and varies $\epsilon_e$. The electron spectral index is fixed to $p=2.3$. All radio constraints are plotted in solid curves.  The X-ray constraint is plotted as the red dotted curve for $\epsilon_e = 0.1$ and $\epsilon_B=0.01$ for comparison.} 
\label{fig:fit}
\end{center}
\end{figure*}

The observational flux is essentially determined by 5 parameters: the redshift of the source $z$ (which would define the blastwave energy $E$ if one assumes that $E$ is comparable to the $\gamma$-ray energy $E_\gamma$, which can be derived from the $\gamma$-ray fluence), 
the ambient medium particle number density $n$, the shock microphysics parameters including $\epsilon_e$, $\epsilon_B$, and the electron spectral index $p$. 
Several statistical works on these parameters have been carried out through modeling individual GRBs, either for the late-time \citep{panaitescu02,yost03} or early-time \citep{liang13,japelj14} broadband afterglow data. Some useful information, such as the typical values or distribution ranges for these parameters, has been collected. For instance, the distribution of $p$ is likely Gaussian ranging from 2 to 3.5, with a typical value 2.5 \citep[e.g.][]{liang13,wang15}. The distribution of $\epsilon_B$ is wide, ranging from $10^{-8}$ to $10^{-1}$ \citep{panaitescu02,santana14,wang15}. The distribution of $\epsilon_e$ is narrow, from $\sim 0.1$ \citep{panaitescu02} to $\sim 0.01$ \citep{gao15}.

\section{Parameter constraints from the radio afterglow upper limit of FRB 131104/Swift J0644.5-5111}

\cite{shannon16} reported an upper limit of 70 $\mu \rm Jy$ and 100 $\mu \rm Jy$ at 5.5 GHz and 7.5 GHz, respectively, in the joint FRB/GRB error circle from 3 days to 2.5 years after FRB 131104 for 25 observational epochs. 
The radio afterglow detection rate of Swift GRBs is about 30\% \citep{chandra12}. Since Swift J0644.5-5111 is a sub-threshold Swift GRB, the non-detection of a radio afterglow is not particularly surprising. On the other hand, the 70 $\mu \rm Jy$ and 100 $\mu \rm Jy$ upper limits are deeper than most GRB radio afterglow upper limits (Fig.5 of \cite{chandra12}) which compensates the low fluence of Swift J0644.5-5111 compared with the majority of Swift GRBs. In any case, the upper limit can pose interesting constraints on the model parameters, e.g. the ambient number density $n$ if one takes the nominal redshift $z = 0.55$, or the redshift $z$ if one takes a typical ISM density $n = (0.1-1)~{\rm cm}^{-3}$.

We first adopt a set of typical values for the microphysics shock parameters ($\epsilon_e$, $\epsilon_B$, and $p$), and then evaluate the excluded region by the upper limit in the 2-dimensional $(z, n)$ parameter space, with $z$ ranging from 0.001 to 0.90, and $n$ ranging from $10^{-4}~\rm{cm^{-3}}$ to $1~\rm{cm^{-3}}$.  
For each pair of $z$ and $n$, all five model parameters are available, and one can calculate the values of characteristic synchrotron frequencies $\nu_m$ and $\nu_a$ at $t=3~\rm days$. For $\nu_{\rm obs}>\rm \max(\nu_a,\nu_m)|_{t=3~\rm days}$, the lightcurve is already in the decaying phase at $t=3~\rm days$. One can require the observational upper flux is higher than $F_\nu$ at $t=3~\rm days$ to constrain parameters in the $(z,n)$ plane. For $\nu_{\rm obs} \leq \rm \max(\nu_a,\nu_m)|_{t=3~\rm days}$, the radio afterglow would be still rising at 3 days, and one should require $F_{\rm \nu,max}$ to be below the observational upper limit to constrain parameters. The rejected parameter spaces in the $(z,n)$ plane for different values of $\epsilon_B$, $\epsilon_e$ are presented. To achieve the most conservative constraints, we use 70 $\mu$Jy at 5.5 GHz to constrain the parameters.

For the commonly used microphysics shock parameters (e.g., $\epsilon_e=0.1$, $\epsilon_B=0.01$ and $p=2.3$, red curve in Fig.1), the radio upper limit for FRB 131104/Swift J0644.5-5111 system could indeed place stringent constraints on the source related parameters $z$ and $n$. The near-flat segment at $z > 0.07$ corresponds to the $\nu_{\rm obs} \leq \rm \max(\nu_a,\nu_m)|_{t=3~\rm days}$ regime, and the curved segment at $z < 0.07$ corresponds to the $\nu_{\rm obs} > \rm \max(\nu_a,\nu_m)|_{t=3~\rm days}$ regime. One can see that if the redshift value is of the order of the one inferred from the DM value (i.e., $z>0.1$), the ambient medium number density should be below $\sim 10^{-3}~\rm{cm^{-3}}$. This disfavors the long GRB models that invoke a massive star origin of the GRB, but is consistent with that inferred from short GRB observations and the theoretical expectations for a compact binary merger environment \citep{berger14}. On the other hand, if one believes that the FRB ambient density is that of a typical ISM ($n= (0.1-1)~{\rm cm}^{-3}$), then a very small redshift is required, i.e. $z\ll0.1$.  This requires that the observed DM is mostly contributed from the host galaxy or the immediate environment near the FRB source instead of the inter-galactic medium. 
If this is the case, one should be cautious to apply FRBs as cosmological probes. 

Besides radio data, the follow-up observations of Swift and VLT also provide X-ray and optical upper limits on the afterglow of FRB 131104/Swift J0644.5-5111, i.e. $4\times 10^{-14} \rm{erg~cm^{-2}~s^{-1}}$ in the X-ray band and 4.5 $\mu \rm Jy$ in the optical band \citep{delaunay16}. The constraints from the X-rays are more stringent, which we also plot as dotted red curve in Fig.1. We find that the constraint on the ambient density is mainly contributed by the radio upper limit at lower redshifts, but the X-ray upper limit places a more stringent constraint in the nominal redshift range inferred by DM. For example, at $z\sim0.55$ the constraint on density placed by X-rays is up to a factor of 2 more stringent than that placed by the radio data (Fig.1). \cite{murase16} and \cite{dai16b} reached a similar conclusion independently.

In view of the wide range of microphysics parameters observed in GRB afterglows, we also test how they impact on the results. For $p$, we test three different values, 2.3, 2.7, 3, and find that the results are barely affected. However, both $\epsilon_B$ and $\epsilon_e$ have significant effects on the constraints. Fixing $\epsilon_e=0.1$ (Fig.1 left panel), the excluded region greatly shrinks as $\epsilon_B$ drops. A long GRB environment is allowed if $\epsilon_B$ is as low as $10^{-4}$.  Similarly, fixing $\epsilon_B=0.01$, the allowed region is also greatly enlarged if $\epsilon_e$ is as small as 0.01.

\section{Conclusion and Discussion}

We have investigated the implications of the upper limit of the radio afterglow of FRB 131104/Swift J0644.5-5111 system reported by \cite{shannon16}. We find that for the nominal redshift values inferred from DM and for typical shock microphysics parameters, $\epsilon_e = 0.1$, $\epsilon_B = 0.01$, a long GRB environment is disfavored but a low-density medium $n<10^{-3}~{\rm cm}^{-3}$ typical for short GRBs of a compact star merger origin is allowed. \cite{murase16} and \cite{dai16b} also discussed the constraints on the ambient medium from radio  and X-ray/optical data, respectively, and similar conclusions have been obtained. If one invokes a typical ISM medium density, then the distance of the FRB is much closer, i.e. $z\ll0.1$, which requires that the large DM is mostly contributed from the FRB host or the immediate environment of the FRB. On the other hand, if one allows $\epsilon_e$ and $\epsilon_B$ to be smaller values, as inferred in some GRB afterglows, the constraints from the upper limit become very loose, so that both long GRB and short GRB environments can be allowed.

Given the loose constraint from the radio afterglow upper limit, the potential FRB 131104/Swift J0644.5-5111 association proposed by \cite{delaunay16} remains plausible. The relative timing of the FRB and the GRB is, however, not easy to interpret within the available models \citep[see also][]{murase16}. Within the supra-massive magnetar collapsing model, the expected FRB would be at the end of the plateau \citep{zhang14}. This is inconsistent with the observation, which shows an extended long GRB following the FRB. Alternatively, one may attribute the FRB to the result of a NS-NS merger, either via synchronization of the magnetosphere \citep{totani13} or pre-merger electromagnetic activity due to a charged member \citep{zhang16a} or the unipolar effect \citep{wang16}. However, such a merger would produce a short GRB rather than a long one. One possible solution is to invoke an off-axis geometry for a NS-NS merger system, with the line-of-sight in the so-called free zone where no dynamical ejecta blocks the line of sight (so that the FRB can reach the observer), but the short GRB jet is missed \citep{zhang13,sun16}. Since NSs are naturally charged, the NS-NS merger may produce an FRB via the mechanism outlined in \cite{zhang16a}. The merger leaves behind a stable magnetar with a spin down time scale of the order of $T_{90} \sim 377$ s, during which the dissipation of the magnetar wind powers the observed long GRB. The luminosity and total energy of the long GRB are consistent with having such a magnetar as the central engine. In any case, the FRB/GRB associations should be rare. Future observations of more FRB/GRB coincident events would allow one to confirm FRB/GRB associations and to test the physical models of these associations.

\acknowledgments
We thank the referee for helpful comments. This work is supported by the National Basic Research Program (`973' Program) of China (grants 2014CB845800), the National Natural Science Foundation of China under grants 11543005, 11603003, 11633001.

{}

\end{document}